\documentclass[%
 reprint,
superscriptaddress,
longbibliography,
 amsmath,amssymb,
 aps,
prl,
]{revtex4-1}
\setcitestyle{numbers,square}

\usepackage{braket}
\usepackage{graphicx}% Include figure files
\usepackage{dcolumn}% Align table columns on decimal point
\usepackage{bm}% bold math
\usepackage[normalem]{ulem} % strikethrough
\usepackage{color}

\usepackage{hyperref}% add hypertext capabilities
\hypersetup{
    colorlinks,%
    citecolor=blue,%
    linkcolor=blue,%
    urlcolor=blue
}

\newcommand{\fp}{\it first-principles}
\newcommand{\mnx}{$M_2NX_3$}
\newcommand{\tmd}{$MX_2$}

\begin{document}

%\title{$M_2NX_3$ a class of Kane-Mele topological materials}
\title{Jacutingaite-family: a class of topological materials}

\author{F. Crasto de Lima} 
\email{felipe.lima@lnnano.cnpem.br}
\affiliation{Brazilian Nanotechnology National Laboratory CNPEM, \\ C.P. 6192, 13083-970, Campinas, SP, Brazil}

\author{R. H. Miwa}
\email{hiroki@ufu.br}
\affiliation{Instituto de F\'isica, Universidade Federal de Uberl\^andia, \\ C.P. 593, 38400-902, Uberl\^andia, MG,  Brazil}

\author{A. Fazzio}
\email{adalberto.fazzio@lnnano.cnpem.br}
\affiliation{Brazilian Nanotechnology National Laboratory CNPEM, \\ C.P. 6192, 13083-970, Campinas, SP, Brazil}

\date{\today}

\begin{abstract}

{Jacutingate, a recently discovered Brazilian naturally occurring mineral, has shown to be the first experimental realization of the Kane-Mele topological model. In this letter we have unveiled a class of materials {$M_2NX_3$} ($M$=Ni, Pt, Pd; $N$=Zn, Cd, Hg; and $X$=S, Se, Te), sharing jacutingaite's key features, i.e., high stability, and topological phase. By employing {\fp} calculations we extensively characterize the energetic stability of this class while showing a common occurrence of the Kane-Mele topological phase. Here we found Pt-based materials surpassing jacutingaite's impressive topological gap and lower exfoliation barrier while retaining its stability.}

\end{abstract}

\maketitle

%%\section{Introduction}

The spin-orbit coupling (SOC) in condensed matter is accountable for the rise of many phenomena. For instance, topological phases \cite{RMPhasan2010}, Rashba states \cite{MATTERcarlos2020}, exotic spin-textures \cite{NATCOMMtao2018}, magnetic anisotropy \cite{COMMPHYSliang2018}, spin-orbit torque transfer \cite{NATCOMMjiang2019} to cite a few. Since the successful synthesis of graphene\,\cite{novoselovScience2004}, 2D materials with sizeable SOC have been pursued as a platform for designing new  electronic/spintronic devices\,\cite{NPJahn2020, JPCMpremasiri2019, RMPavsar2020}, as well as the discovery and confirmation of new physical phenomena. For instance, two-dimensional hexagonal lattices and SOC are the key ingredients to the manifestation of the quantum spin Hall (QSH) state as predicted by Kane and Mele\,\cite{PRLkane2005}.

Indeed, recent research addressing topological phases in low dimensional systems has boosted the search for new two dimensional materials with large SOC, for instance, the 2D MXenes \cite{NLsi2016, ACSNANOfrey2018}, and  Bi-based systems \cite{JMCCji2017, SCIENCEreis2017, PRBsingh2017}. Other strategies have been proposed based on  the incorporation of adatoms\,\cite{weeks2011engineering, PRLzhang2012, PRBacosta2014, ACSNANOklimovskikh2017, RSCADVafzal2019, PRBwakamura2019} and proximity effects\,\cite{qiao2014quantum, wang2015proximity, mendes2015spin} in order to strengthen the spin-orbit effects. However, most of these routes face a sort of difficulties, namely  (i) structural metastability, (ii) stability against environment conditions (for instance, oxidation processes),  (iii) difficulties for the experimental realization, (iv) lost of 2D character due to the electronic interaction with a given substrate. Thus, new 2D materials with large SOC, that overcomes the hindrances above, are still quite desirable.

\begin{figure}[h!]
\includegraphics[width=\columnwidth]{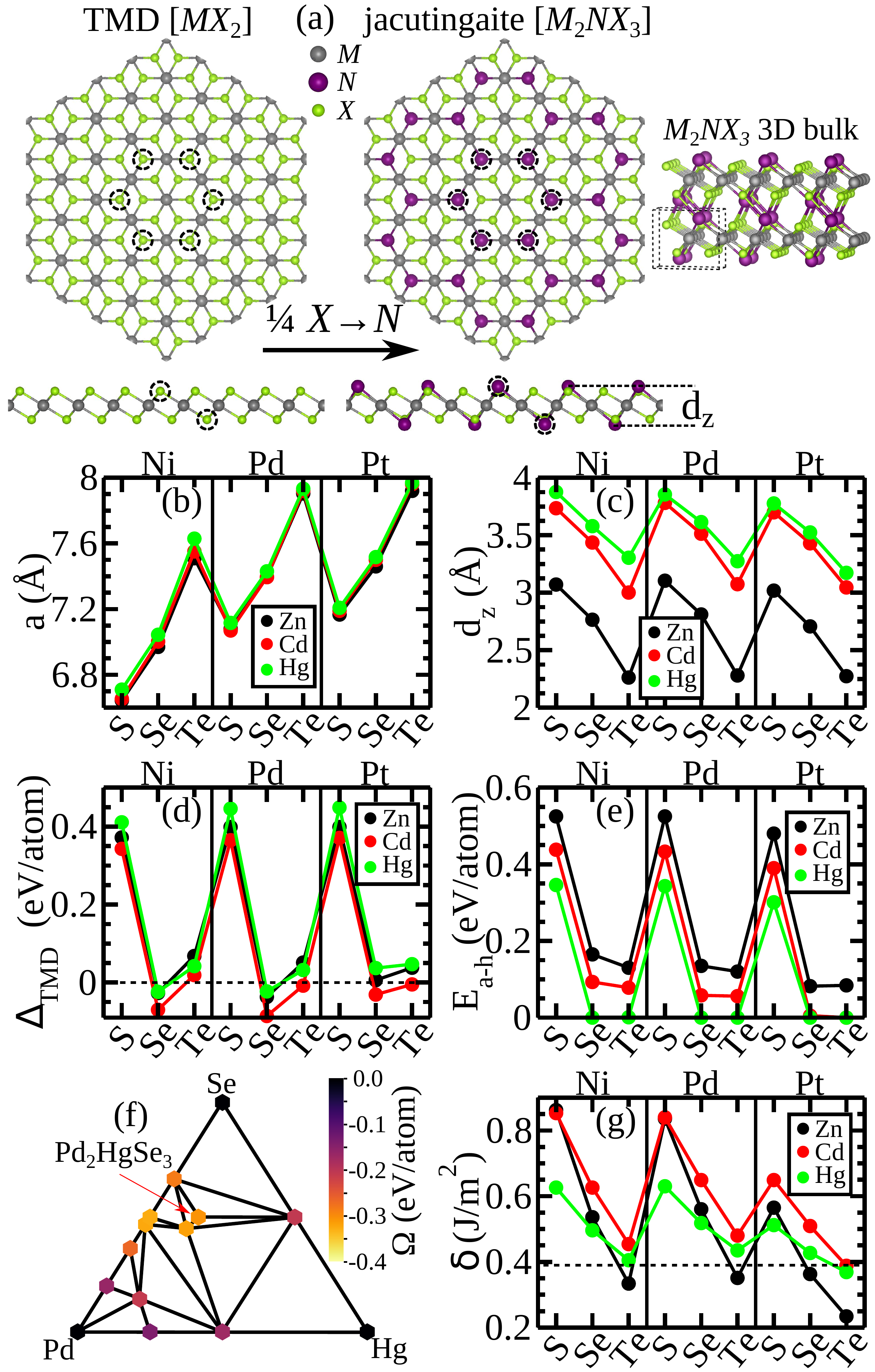}
\caption{\label{1L-par} (a) {2D} TMD ({\tmd}) and {2D/3D} {\mnx} atomic structure, (b) lattice parameter, (c) bucking distance, (d) formation energy comparison between the TMD and {\mnx}, (e) ternary $MNX$ energy above hull for {\mnx}, (f) convex hull for Pd$_2$HgSe$_3$ and (g) cleavage energy with the dashed line indicating the graphite cleavage barrier.}
\end{figure}

Jacutingaite (Pt$_2$HgSe$_3$), a naturally occurring mineral, discovered in Brazil by  Cabral {\it et al.}\,\cite{cabral2008platinum} in 2008, is gaining attention in the last few years. The large SOC and the honeycomb structure of Hg atoms make jacutingaite, in its monolayer form, a place for the manifestation of the Kane-Mele topological phase\,\cite{PRLmarrazzo2018}; whereas in its bulk form it has been predicted to host dual topology, being a weak topological insulator and a topological crystaline insulator\,\cite{PRMfacio2019, PRBghos2019, PRRmarrazzo2020}. Both topological phases, namely Kane-Mele in monolayer and dual in bulk Pt$_2$HgSe$_3$, have been experimentally observed through  scanning tunneling microscopy (STM)\,\cite{NLkandrai2020} and angle-resolved photoemission (ARPES)\,\cite{PRLcucchi2020} measurements. Its atomic structure [Fig.~\ref{1L-par}(a)] can be viewed as the transition metal dichalcogenide (TMD) PtSe$_2$ with 1T structural phase, where  $1/4$ of the chalcogenides are replaced by Hg.  Besides being a natural occurring mineral, which indicates its stability in geological pressures and temperatures, Pt$_2$HgSe$_3$ has been experimentally synthesized\,\cite{CanMinanna2012, JRSlonguinhos2020, NLkandrai2020, PRLcucchi2020}, as well as its counterpart Pd based jacutingaite like structure, Pd$_2$HgSe$_3$\,\cite{PDlaufek2017}. Those findings suggest that further combinations of 1T TMD host with  high SOC elements may result in novel quantum spin Hall insulators based on jacutingaite like 2D structures.

In this letter, we unveil a large class of highly stable materials hosting QSH and Z$_2$ metallic phases based on the Kane-Mele model. We employ density functional theory (DFT) simulations \footnote{See supplemental materials at http://XXX for the computational details citing references \cite{vasp1, PBE, JCPheyd2003, JCPheyd2006, PRBmonkhorst1976, PRBblochl1994, VDW-DF2, PRBchaput2011, SMtogo2015}} of the structural and electronic properties of {\mnx} compounds with $M$= Ni, Pd, Pt; $N$= Zn, Cd, Hg; and $X$=S, Se, Te. Throughout the extensive analysis of phase diagrams based on $MNX$ compounds, we present a well-grounded prediction of the  energetic stability of  jacutingaite like {\mnx} structures. Electronic structure calculations reveal the emergence of topological phases, with SOC induced (non-trivial) band gaps surpassing the one of jacutingaite. Further topological characterization reveals trivial\,$\rightarrow$\,non-trivial transition as a function of the chemical composition, $X$\,=\,S\,$\rightarrow$\,Te. Finally, addressing the design of nanodevices, we have examined some key features of these  jacutingaite like structures, namely the stability/behavior of the (i) topological gap as a function of the mechanical strain, and the (ii) work function for the different {\mnx} combination.

%%%%%%FIG

%%%%%%%%%%%%%%%%%%%%%%%%%%%%%%%
%%Structural Characterization%%
%%%%%%%%%%%%%%%%%%%%%%%%%%%%%%%

As shown in Fig.~\ref{1L-par}(a), the {\mnx} jacutingaite like  share the same backbone geometry of the 1T TMDs ({\tmd}), where the chalcogenide atoms ($X$) are partially replaced by transition metals ($N$), {\tmd}\,$\rightarrow$\,{\mnx}, resulting in buckled $N$--$M$--$N$ bonds. The $N$ atoms form triangular lattices  on the opposite sides of the {\tmd} host, which in turn are rotated by 60 degrees with respect to each other, giving rise to a buckled hexagonal lattice. At the equilibrium geometry, the lattice constants of the {\mnx} structures are practically independent of the transition metal, i.e. nearly the same  as those of the hosts ({\tmd}).  For instance, the equilibrium lattice constants of Pt$_2N$Se$_3$, for $N$ = Zn, Cd, and Hg, differ  by less than  0.9\%, compared with that of 1T PtSe$_2$. Such independence is due to the $N$--$M$--$N$ buckled structure [Fig.\,\ref{1L-par}(a)] acting as a source of strain relief induced by the foreign ($N$) atom. As shown in Fig.\,\ref{1L-par}(c), the vertical buckling (d$_z$) of the $N$--$M$--$N$ bonds present larger (lower) values for $X$=S (Te). 

The energetic stability of the jacutingaite like structures can be examined by comparing the formation energy of {\mnx} with the one of its respective (energetically stable) {\tmd} host, $\Delta_{\rm TMD} = \Omega[{M}_2{NX}_3] - \Omega[{M}{X}_2]$, Fig.~\ref{1L-par}(d). Here the formation energy is given by a total energy difference between the compound $x$ final system ($E[{x}]$), and the upper limit of the chemical potentials of its isolated compounds ($\mu^{\rm bulk}$), namely, 
$$
\Omega[x] = E[{x}] - \sum_i n_i \mu_i^{\rm bulk},
$$  
where $n_i$ indicates its number of atoms of the specie $i$=$M$, $N$, and $X$. Our $\Delta_{\rm TMD}$ results reveal that the jacutingaite like structures is quite likely to occur for $X$=Se and Te. Here, we found negative values of $\Delta_{\rm TMD}$ for the former, while for $X$=Te it increases by less than 0.1\,eV/atom, Fig.~\ref{1L-par}(d). Meanwhile, for $X$=S the {\mnx} structure is less stable than it hosts by about 0.4\,eV/atom.

Further structural stability of the jacutingaite like {\mnx} structures has been examined through conver energy hull analysis,  comparing their formation energies ($\Omega$) with other $MNX$ ternary phases extracted from the Materials Project database \cite{CMong2008, ECong2010, APLMjain2013}. We found {\mnx} compounds being a node point in the convex hull (zero energy above convex hull, $E_\text{\rm a-h} = 0.0$\,eV/atom \cite{SM}) showing its experimental stability, Fig.~\ref{1L-par}(e). For instance, in Fig.~\ref{1L-par}(f), Pd$_2$HgSe$_3$ lies in a convex node with a formation energy of $-0.18$\,eV/atom. Additionally, all $M_2$HgSe$_3$ ($M$=Ni, Pd, Pt) have $E_\text{\rm a-h} = 0.0$, as well as  Pd$_2$HgTe$_3$, Pt$_2$CdTe$_3$ and Pt$_2$HgTe$_3$. For the Se and Te based materials that have non-zero energy above hull we found $E_\text{\rm a-h} < 0.18$\,eV/atom, which indicate its high stability \cite{ACSAMIschleder2020}. For instance, taking Pt$_2$ZnTe$_3$ as a case of study [$E_\text{\rm a-h} = 0.08$\,eV/atom], we have calculated its monolayer phonon dispersion \cite{SM}, where its dynamical stability was confirmed by the absence of negative frequencies. Additionally, for the higher $E_\text{\rm a-h}$ systems, $X$=S based compounds, their negative values of  formation energies, $\Omega<0$ \cite{SM}, indicates that they can be experimentally stabilized throughout specific synthesis routes and/or substrate support. {Although the SOC has a stabilizing role in the jacutingaite phonon dispersion \cite{PRLmarrazzo2018}, we see that it changes the formation energy by $\sim 7$\,meV/atoms which do not change our conclusions.}

The cleavage energy ($\delta$)\,\footnote{$\delta = (E_{bulk} - E_{ML})/A$, with $E_{bulk}$ ($E_{ML}$) the total energy of the bulk (monolayer) system, and $A$ monolayer the unity cell area.} is another important structural information for the top-down synthesis of 2D systems. We found that the {\mnx} bulk phase presents cleavage energies in the range of other experimentally exfoliated materials \cite{SRchoudhary2017}. For instance, jacutingaite has a cleavage energy ($\delta = 0.46$\,J/m$^2$) comparable with that of graphene exfoliated from graphite, $\delta = 0.39$\,J/m$^2$ [dashed line in Fig.\,\ref{1L-par}(g)]\,\cite{NATCOMMwang2015}. Comparing the calculated cleavage energy and the vertical buckling of the $N$--$M$--$N$ bonds [$d_z$ in Fig.\,\,\ref{1L-par}(c)] it is noticeable that, (i) for a given {transition metal pair $M$-$N$}  the $\delta$ is proportional to  $d_{z}$, being larger for $X$=S and lower for $X$=Te, it is in agreement   with Ref.\,\cite{CanMinanna2012}, where the authors verified  that the $N$ atoms are responsible for the inter-plane bound of the {\mnx} system{, as shown in Fig.~\ref{1L-par}(a) {\mnx} 3D structure. Indeed, taking the Cd and Hg systems, which have a similar bucking distance, the former presents a stronger interlayer bond ruled by the bonding energy of Cd-$X$ being $\sim 60\%$ greater than Hg-$X$. By contrast, for Zn, given its lower buckling distance, an interplay between the vdW interaction (of the TMD host) and the Zn interlayer chemical bond is present. Here the later dominating for lower $d_z$ and leading to weaker interlayer bond.} (ii) For $X$=Te and $N$=Zn, the cleavage energy are always lower than that of graphene, where (iii)  the cleavage energy of Pt$_2$HgTe$_3$ is lower when compared with its counterpart (Pt$_2$HgSe$_3$) jacutingaite.  

%%%%%%%%%%%%%%%%%%%%%%%%%%%%%%%
%%Electronic Characterization%%
%%%%%%%%%%%%%%%%%%%%%%%%%%%%%%%

%%%%%%FIG
\begin{figure}[h!]
\includegraphics[width=\columnwidth]{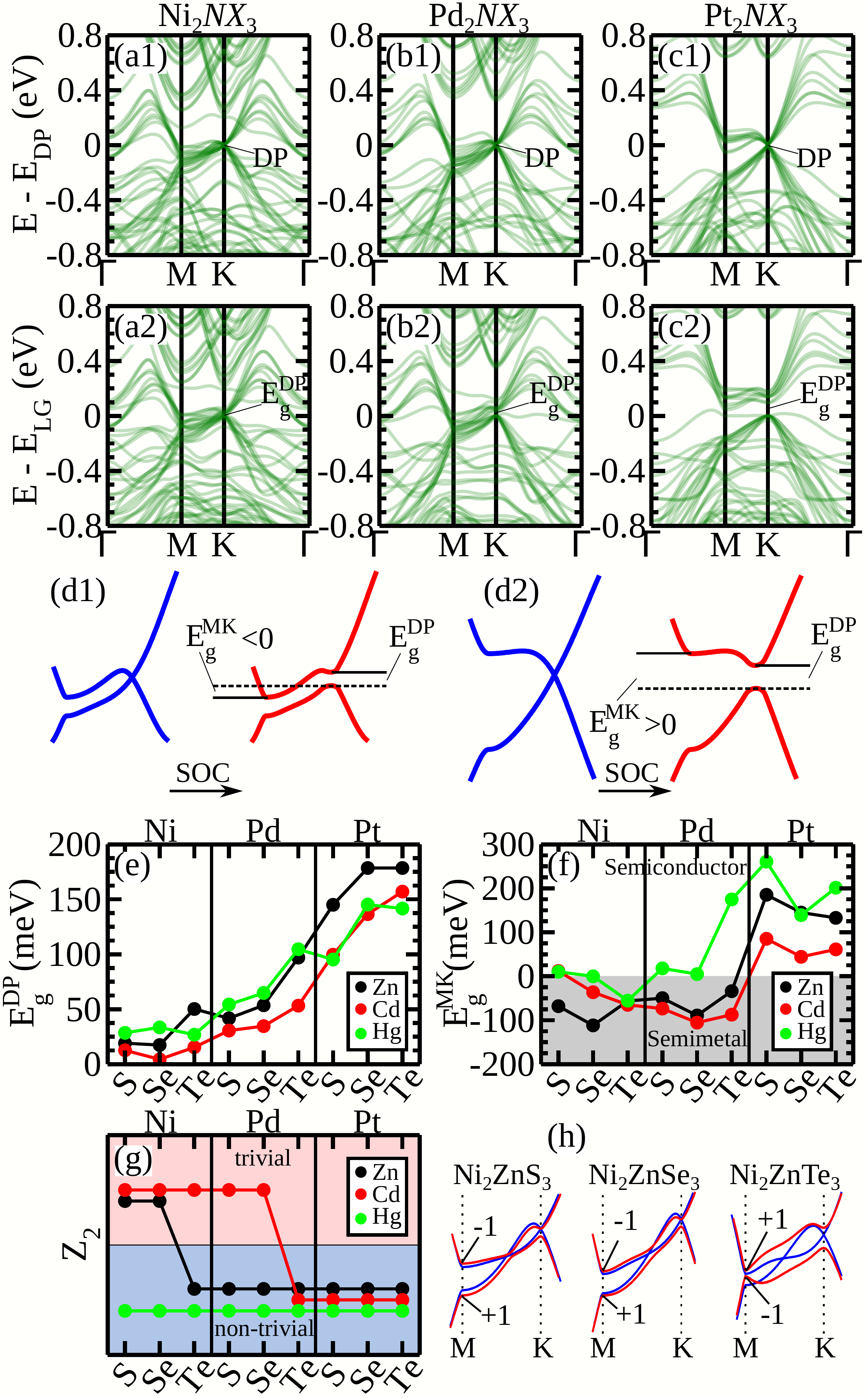}
\caption{\label{soc-gap} Superposition of band structures for the (a) {Ni$_2NX_3$}, (b) {Pd$_2NX_3$} and (c) {Pt$_2NX_3$} systems, (a1)-(c1) without SOC and energy in relation to the DP ($E_{DP}$) and (a2)-(c2) with SOC and energy in relation to the K point gap's lower band ($E_{LG}$). (d) Depiction of the Dirac bands and SOC effect in the (d1) downward band bending at M and (d2) graphene-like. (e) Dirac point SOC energy gap ($E_g^{DP}$); (f) global energy gap between M and K points ($E^{MK}_g$); (g) topological invariant; and (h) topological transition for {Ni$_2$Zn$X_3$} with blue (red) lines indicating the states without (with) SOC.}
\end{figure}

Once we have shown the feasibility of the energetically stable   counterparts of jacutingaite, here we will focus on the electronic properties and topological phases of single layer {\mnx} systems. As shown in Fig.~\ref{soc-gap}(a)-(c), we found the emergence of Dirac cones ruled by the hexagonal $N$--$M$--$N$ buckled lattice. The projection of the energy bands, near the Fermi level, reveals that the Dirac cones are mostly composed  by the transition metal $N$($s$) orbitals  hybridized with the host $M$($d$), orbitals {\it viz.}: Ni(3$d$), Pd(4$d$), and Pt(5$d$). It is noticeable that, for a given host transition metal $M$, the electronic band structures share nearly the same features around the Fermi level. For instance, in Fig.\,\ref{soc-gap}(a) we present the superposition of the electronic band structures of Ni$_2NX_3$ with $N$=Zn, Cd, and Hg, and $X$=S, Se, and Te; similarly for Pd$_2NX_3$ and Pt$_2NX_3$, as shown in Figs.\,\ref{soc-gap}(b) and (c). {Such figure, where each material contributes with a translucent set of lines (band structure), allow us to identify similar features in the compounds characterized by darker regions (more details at supplemental material\,\cite{SM}).} In the  absence of SOC, the linear dispersion of the energy bands at the K and K$^\prime$  gives rise to the  Dirac points (DPs) indicated as DP in Figs.~\ref{soc-gap} panels (a1)-(c1). Whereas,  by turning on the SOC contribution, we find  energy gaps taking place at the DPs [$E_g^{\rm DP}$ in Fig.~\ref{soc-gap} panels (a2)-(c2)]. The SOC in the system are most given by the $M$ atoms \cite{PRLmarrazzo2018}, and its strength is a quantitative indication of the stability of topological states. Here, it is worth to highlight that SOC induced energy gaps at the DP [Fig~\ref{soc-gap}(e)] are larger in Pt$_2$Zn$X_3$ compared with the ones of the other Pt$_2NX_3$ systems. For instance, Pt$_2$Zn$X_3$ (with $X$ = Se or Te) presents $E_{\rm g}^{\rm DP} \approx 178$\,meV, being larger by 34\,meV (23\%) compared with that of jacutingaite and its counterpart Pt$_2$HgTe$_3$, both systems present $E_{\rm g}^{\rm DP}=145$\,meV. In particular, these findings can be understood by comparing the equilibrium geometries of the $N$--Pt--$N$ buckled hexagonal lattice; namely Zn--Pt--Zn presents lower values of vertical buckling and Zn--Pt equilibrium bond length ($d_{MN}=2.55$\,\AA) compared with the ones of Hg--Pt--Hg, $d_{MN}=2.79$\,\AA, strengthening the Pt contribution to the Zn($s$) Dirac bands. It is worth pointing out that larger values of $E_{\rm g}^{\rm DP}$ in Pt$_2$Zn$X_3$ have been maintained even upon the use of hybrid functionals (HSE). Here we found $E_{g}^{DP} = 232$ and $242$\,meV for $X$ = Se and Te,  both larger than that obtained for  jacutingaite, 222\,meV (in Ref.\,\cite{PRBwu2019} the authors obtained 218\,meV using the same calculation approach). 

Besides the energy gap induced by the SOC at the K/K' points,  downward bending of the upper Dirac band along the K--M direction leads to lower values of global gaps, and eventually resulting in semimetallic systems.  In Fig.~\ref{soc-gap}(f) we show the energy difference between the lower point of the upper Dirac band at the M-point and the top of the Dirac valence band at K-point [$E_g^{\rm MK}$ in Figs.~\ref{soc-gap}(d1)-(d2)]. Negative values of $E_{g}^{\rm MK}$, for $M$ = Ni and Pd systems, indicate that they are semimetallic. For $M$ = Pt{, the SOC strength always overcome the downward band bending, where} the semiconducting character has been preserved. Focusing on the Pt$_2N$Te$_3$ systems, we found indirect energy band-gaps of 133 and 61\,meV for $N$ = Zn and Hg, respectively, while Pt$_2$HgTe$_3$ presents a direct energy gap  ($E_{\rm g}^{\rm DP} < E_{\rm g}^{\rm MK}$) of 142\,meV.

%%%Higher SOC-GAP (meV)
%Pt2ZnS3  = 145.0 ; Pt2CdS3  =  99.6 ; Pt2HgS3  =  95.1
%Pt2ZnSe3 = 178.4 ; Pt2CdSe3 = 136.5 ; Pt2HgSe3 = 145.2
%Pt2ZnTe3 = 178.5 ; Pt2CdTe3 = 157.0 ; Pt2HgTe3 = 141.6

To characterize the topological phase of the {\mnx} systems we have computed the Z$_2$\,\cite{PRLkane22005} invariant by analyzing the parity of each band at the time-reverse invariant momenta (TRIM) and considering all bands bellow the upper Dirac bands fully occupied\,\cite{PRBfu2007}. It is worth noting that the presence of  semimetallic  bands does not necessarily rule out the (possible) emergence of topologically non-trivial phases, characterizing a Z$_2$-metallic phase\,\cite{PRBzhao2014}. In this case, the edge states are no longer protected against backscattering processes. Our results, summarized in Fig.~\ref{soc-gap}(g), reveal that while all Hg compounds present  a non-trivial topological phase,  Ni$_2$Zn$X_3$ and Pd$_2$Cd$X_3$ systems present a trivial\,$\rightarrow$\,non-trivial topological transition for $X$ = S\,$\rightarrow$\,Te. In these systems, there is an extra crossing of the Dirac bands (without SOC) along the M--K path [blue lines in Fig.~\ref{soc-gap}(h)], which causes a parity inversion from the original graphene-like bands. Such an inversion is overcome by the SOC, changing the parity order at the TRIM, for $X$ = Te leading to the Z$_2$ non-trivial topological phase, whereas in Ni$_2$Zn$X_3$  the strength of the SOC  is not enough to re-invert this extra crossing.

In addition to the topological phase control (trivial\,$\leftrightarrow$\,non-trivial), as a function of the chemical composition in {\mnx} jacutingaite like structures; other properties worth to be examined in order to achieve technological applications, as well as the integration with other 2D materials. For instance, the behavior of the  SOC (non-trivial) energy gap as a function of the mechanical strain.  Here, we found that the topological phase of the {\mnx} jacutingaite like systems are robust against biaxial strain. In particular, for Pt$_2$ZnSe$_3$,  there is an increase of  the SOC induced non-trivial energy band-gaps, $E^{\rm DP}_{\rm g}$, from  178 to 185\,meV upon a compressive strain of 2\,\%. It reaches to $\sim$194\,meV in Pt$_2$ZnTe$_3$ compressed by  2.5\%. Another key feature, in order to perform electronic engineering based on {\mnx} 2D heterostructures\,\cite{NATUREgeim2013}, is the  work function ($\Phi$). By examining its dependence with the chemical composition, we found that for a given {\tmd} host,  $\Phi$ varies as a function of the foreign transition  metal $N$. For instance, considering the PtSe$_2$ host, we found $\Phi$ = 4.90\,eV for jacutingaite ($N$=Hg), and it increases to about 4.70\,eV in Pt$_2$ZnSe$_3$. Meanwhile, in Pt$_2N$Te$_3$ $\Phi$ reduces to 4.73 and 4.43\,eV for $N$ = Hg and Zn, respectively. Further details of the SOC induced bandgap as a function of biaxial strain and {\mnx} work functions are shown in the supplemental material\,\cite{SM}.

%%monolayer WF
%Pt2HgS3  = 5.20 eV ; Pd2HgS3  = 5.23 eV ; Ni2HgS3  = 4.97 eV
%Pt2CdS3  = 4.43 eV ; Pd2CdS3  = 4.68 eV ; Ni2CdS3  = 4.38 eV
%Pt2ZnS3  = 4.66 eV ; Pd2ZnS3  = 4.46 eV ; Ni2ZnS3  = 4.69 eV
%Pt2HgSe3 = 4.91 eV ; Pd2HgSe3 = 5.06 eV ; Ni2HgSe3 = 4.88 eV
%Pt2CdSe3 = 4.40 eV ; Pd2CdSe3 = 4.50 eV ; Ni2CdSe3 = 4.37 eV
%Pt2ZnSe3 = 4.54 eV ; Pd2ZnSe3 = 4.68 eV ; Ni2ZnSe3 = 4.60 eV
%Pt2HgTe3 = 4.73 eV ; Pd2HgTe3 = 4.80 eV ; Ni2HgTe3 = 4.57 eV
%Pt2CdTe3 = 4.38 eV ; Pd2CdTe3 = 4.46 eV ; Ni2CdTe3 = 4.33 eV
%Pt2ZnTe3 = 4.42 eV ; Pd2ZnTe3 = 4.56 eV ; Ni2ZnTe3 = 4.50 eV

%%\section{Conclusion}

In conclusion, we have shown that beyond the jacutingaite, which is a naturally occurring mineral,  there are other jacutingaite like ({\mnx}) structures that also host the Kane-Mele QSH phase. The energetic stability of these {\mnx} jacutingaite counterparts was thoroughly investigated, based on formation energy calculations combined with  the recently convex hull analysis. We found that the {\mnx} structures are quite stable for chalcogenide ($X$) Se and Te atoms. Further calculations of the cleavage energies show that these systems can be isolated in monolayers by commonly used exfoliation methods. The topological phases were characterized upon the calculation of Z$_2$ and Z$_2$-metallic invariants, where we found that  for $N$ = Hg the non-trivial phase has been preserved for any  combination between the transition metal $M$ (= Ni, Pd, and Pt) and chalcogenide atom $X$ (= S, Se, and Te). In contrast, for the other foreign transition metals, $N$ = Zn and Cd, the topological phase depends on the combination of the host ({\tmd}) atoms. Finally,  we found that the Zn compounds, Pt$_2$ZnSe$_3$ and Pt$_2$ZnTe$_3$, present larger values of topological bandgap compared with that of jacutingaite (Pt$_2$HgSe$_3$), while by changing the host element, $X$ = Se\,$\rightarrow$\,Te, the bandgap is practically the same, however, it becomes direct. Nevertheless, the unveiled class of materials introduces a family of compounds hosting the Kane-Mele topological phase, which can be of interest in devices engineering.

\begin{acknowledgments}

The authors acknowledge financial support from the Brazilian agencies FAPESP (grants 19/20857-0 and 17/02317-2), CNPq, and FAPEMIG, and the CENAPAD-SP and Laborat\'{o}rio Nacional de Computa\c{c}\~{a}o Cient\'{i}fica (LNCC-SCAFMat2) for computer time.

\end{acknowledgments}

%\appendix

\bibliography{bib}% Produces the bibliography via BibTeX.

\end{document}